\documentclass{article}
\usepackage[numbers,sort&compress,square]{natbib}
\usepackage[lang=en]{jabbrv}
\usepackage{dcase2021,graphicx,url,times}
\usepackage{amsmath, amsthm, amssymb, amsfonts}
\usepackage{comment}
\usepackage{booktabs}
\usepackage{siunitx}
\usepackage{tabularx}
\usepackage{bm}
\usepackage{placeins}
\usepackage{cleveref}
\usepackage{lipsum}

\usepackage{etoolbox}           
\renewcommand{\bfseries}{\fontseries{b}\selectfont} 
\robustify\bfseries             
\newrobustcmd{\B}{\bfseries}    
\newcommand{\V}{$\times$}
\newcommand{\X}{$\checkmark$}
\newcommand{\mbf}{\mathbf}

\newcolumntype{C}{>{\centering\arraybackslash}X}
\newcommand{\Fl}{F\textsubscript{1} }

\usepackage{pgfplots}
\pgfplotsset{width=\columnwidth,compat=1.9}

\title{Improving Polyphonic Sound Event Detection on Multichannel Recordings with the Sørensen--Dice Coefficient Loss\\and Transfer Learning}

\name{
    Karn N. Watcharasupat$^{1*}$,
	Thi Ngoc Tho Nguyen$^{1*}
        \thanks{This research was supported by the Singapore Ministry of Education Academic Research Fund Tier-2, under research grant MOE2017-T2-2-060.} \thanks{K. N. Watcharasupat acknowledges the support from the CN Yang Scholars Programme, Nanyang Technological University, Singapore.}\thanks{*Equal contribution.}$,
} \secondlinename{	  
	Ngoc Khanh Nguyen,
	Zhen Jian Lee,
    Douglas L. Jones$^{2}$, 
    Woon Seng Gan$^{1}$
}

\address{
    $^1$School of Electrical and Electronic Engineering, Nanyang Technological University, Singapore.\\          
    $^2$Dept. of Electrical and Computer Engineering, University of Illinois at Urbana-Champaign, IL, USA. \\
    Emails: \{karn001, nguyenth003\}@e.ntu.edu.sg, \{ngockhanh5794, zhenjianlee\}@gmail.com,\\ dl-jones@illinois.edu, ewsgan@ntu.edu.sg
}

\begin{document}

\ninept
\maketitle

\begin{sloppy}

\begin{abstract}
The Sørensen--Dice Coefficient has recently seen rising popularity as a loss function (also known as Dice loss) due to its robustness in tasks where the number of negative samples significantly exceeds that of positive samples, such as semantic segmentation, natural language processing, and sound event detection. Conventional training of polyphonic sound event detection systems with binary cross-entropy loss often results in suboptimal detection performance as the training is often overwhelmed by updates from negative samples. In this paper, we investigated the effect of the Dice loss, intra- and inter-modal transfer learning, data augmentation, and recording formats, on the performance of polyphonic sound event detection systems with multichannel inputs. Our analysis showed that polyphonic sound event detection systems trained with Dice loss consistently outperformed those trained with cross-entropy loss across different training settings and recording formats in terms of \Fl score and error rate. We achieved further performance gains via the use of transfer learning and an appropriate combination of different data augmentation techniques, which mitigate the problem of lacking training data.
\end{abstract}

\begin{keywords}
Polyphonic sound event detection, microphone array, first-order ambisonics, dice loss, deep learning
\end{keywords}

\section{Introduction}
\label{sec:intro}

Sound event detection (SED) is the task of jointly detecting the onset, offset, and class of a sound event. Polyphonic SED refers specifically to the task of such detection for multiple, potentially overlapping sound events simultaneously. In the past decade, there have been significant advances in the applications of deep learning for SED~\cite{Mesaros2021SoundTutorial}. The state-of-the-art SED models have been built from convolutional neural networks (CNN)~\cite{kong2019panns}, convolutional recurrent neural networks (CRNN)~\cite{cakir2017convolutional}, and more recently, Conformer~\cite{Miyazaki2020Conformer-BasedAugmentation}.

Majority of works in SED targeted single-channel inputs, due to the availability of large-scale datasets such as AudioSet~\cite{Gemmeke2017AudioEvents} and FSD50k~\cite{Fonseca2017FreesoundDatasets}, and ease of practical deployment. For multichannel SED (MSED), there are only a few publicly-available labelled datasets, often small-scale, e.g., TUT Sound Events 2016~\cite{Mesaros2016TUTDetection}, and TAU-NIGENS Spatial Sound Events 2021~\cite{politis2021dcasedataset}. In addition, many MSED works naturally focused on multichannel and spatial feature engineering~\cite{adavanne2017sedspatial, Nguyen2020OnDetection} due to the different multichannel formats available. As a result, there is a severe lack of literature concerning the impact of different data augmentation techniques, network architectures, and loss functions on the performance of MSED models that are trained on small datasets. 

In this paper, we present a thorough investigation on the factors affecting the performance of deep polyphonic MSED systems. We demonstrated the effectiveness of using pretrained weights from single-channel model for MSED to tackle the lack of training data. We also investigated the use of dice loss, which has been shown to improve single-channel SED performance~\cite{Imoto2021ImpactPerformance}. Several experiments on the data augmentation techniques, training chunk durations, pretraining modalities, and multichannel audio formats were performed and their results were analyzed and discussed to better inform our understanding of polyphonic MSED systems. The rest of the paper is organized as follows. \Cref{sec:prelim} describes our experiment setups. \Cref{sec:exp} presents the experimental results and discussion. Finally, we conclude the paper in \Cref{sec:concl}. 

\section{Preliminaries}
\label{sec:prelim}

\subsection{Dataset}

In this paper, we used the TAU-NIGENS Spatial Sound Events 2021 (SSE21) dataset~\cite{politis2021dcasedataset}, which provides two four-channel array formats: first-order ambisonics (FOA) and microphone array (MIC). Each format has \num{400}, \num{100}, and \num{100} \num{60}-second recordings for training, validation, and testing, respectively. Both formats were used to train and evaluate the performance of the SED networks. Since SSE21 is a sound event localization and detection (SELD) dataset with directional interferences, we only take the SED labels. Hence, we have strongly-labeled sound events from the $L=12$ target classes and unlabelled out-of-class interferences. 

\subsection{Input features}

Polyphonic \mbox{$C$-channel} audio can generally be modelled in the time-frequency (TF) domain by
\begin{equation}
    \textstyle\mbf{X}[t, f] = \sum_i \mbf{H}_i[t, f]S_i[t, f] + \mbf{N}[t, f]\in\mathbb{C}^C,
\end{equation}
where $S_i[t, f]\in\mathbb{C}$ is the $i^{th}$ sound source, $\mbf{H}_i[t, f]\in\mathbb{C}^{C}$ is the corresponding array response vector from the source to the sensor array, and $\mbf{N}[t, f]\in\mathbb{C}^{C}$ is the noise vector.
This representation is typically achieved by applying the short-time Fourier transform (STFT) on the multichannel time-domain audio data. A Hann window of size $R=1024$ with hop size \num{300} was used for this paper.

For the FOA format, the array response vector is given by
\begin{equation}
    \mbf{H}^\text{(FOA)}_i[t, f]
    = \begin{bmatrix}
    H_{i, \text{w}}[t] \\
    H_{i, \text{y}}[t] \\
    H_{i, \text{z}}[t] \\
    H_{i, \text{x}}[t] 
    \end{bmatrix} = \begin{bmatrix}
    1 \\
    \sin(\phi_i[t]) \cos(\theta_i[t])\\
    \sin(\theta_i[t]) \\
    \cos(\phi_i[t]) \cos(\theta_i[t])
    \end{bmatrix} \in \mathbb{R}^{4},
\end{equation}\nobreak
where $\phi_i[t]$ and $\theta_i[t]$ are the time-dependent azimuth and elevation angles of the $i^{th}$ sound source with respect to the array, respectively.

For the microphone format, SSE21 provides a tetrahedral array of microphones mounted on a spherical baffle, whose response has a very complex analytical expression~\cite[eq. (5)]{politis2020dcasedataset}. From our experience, we have found that a generic far-field array response can usually approximates the theoretical response satisfactorily. The far-field response for the $c^{th}$ channel at time $t$ is given by 
\begin{equation}
    H^\text{(MIC)}_{i,c}[t, f]
    = \exp\left(-\dfrac{\jmath 2\pi f d_{i,c}[t]}{Rv/f_\text{s}}\right) \in \mathbb{C},
\end{equation}
up to some channel-dependent scaling, where $v \approx \SI{343}{\meter\per\second}$ is the speed of sound, and $d_{i,c}[t]$ is the projected difference in distances, in metres, travelled by the $i^{th}$ sound source arriving the $c^{th}$ microphone relative to the reference microphone.

In this paper, \num{128}-bin log-mel spectrograms were used as input features, with cutoff frequencies at \SI{50}{\hertz} and \SI{12}{\kilo\hertz}. To facilitate learning, \textit{z}-score normalization was performed on each channel of the spectrograms along the mel-frequency bin axis. 


\subsection{Data augmentation}\label{ssec:aug}

As with many audio datasets, SSE21 is relatively small, thus requiring data augmentation. In this paper, we experiment with combinations of four data augmentation techniques: mixup with soft label~\cite{zhang2017mixup}, frequency shifting, channel swapping~\cite{Mazzon2019}, and a composite technique of cutout~\cite{DeVries2017ImprovedCutout} and SpecAugment~\cite{park2019specaugment}. 

\subsection{Loss function}

In SED, the number of negative samples often significantly exceeds that of positive samples. Hence, training polyphonic SED systems with binary cross-entropy (BCE) loss often results in suboptimal \Fl score. An alternative loss function based on the Sørensen--Dice Coefficient (SDC)~\cite{Dice1945MeasuresSpecies, Srensen1948ACommons}, called Dice loss, was proposed in~\cite{Milletari2016V-Net:Segmentation}. Variations of the Dice loss have since become popular as loss functions in semantic segmentation~\cite{Sudre2017GeneralisedSegmentations} and natural language processing. A variant was also recently used in single-channel SED~\cite{Imoto2021ImpactPerformance}.

The SDC between two sets $\mathfrak{A}$ and $\mathfrak{B}$ is defined by
\begin{equation}
    \text{SDC}(\mathfrak{A}, \mathfrak{B})
    = {2\lvert\mathfrak{A}\cap\mathfrak{B}\rvert}/{\left(\lvert\mathfrak{A}\rvert+\lvert\mathfrak{B}\rvert\right)}
\end{equation}
where $\lvert\cdot\rvert$ is the set cardinality operator. The SDC is used to gauge the similarity between two sets. In binary classification, the SDC is identical to the F\textsubscript{1} score. Generalizing the SDC to tensors with elements in the unit interval, the cardinality operator can be replaced by the $L_1$-norm, $\|\cdot\|_1$, while the intersection operator can be replaced by the Hadamard multiplication, $\circ$. As such, a differentiable Dice loss for a batch of $N$ predictions is given by
\begin{equation}
    \mathcal{L}_\text{Dice}(\hat{\mbf{Y}}, \mbf{Y})
    = \dfrac{1}{N}\sum_n \left[1 - \dfrac{2\|\widehat{\mbf{Y}}_n \circ  \mbf{Y}_n\|_1}{\|\widehat{\mbf{Y}}_n+\mbf{Y}_n\|_1 + \epsilon}\right]
\end{equation}
where $\widehat{\mbf{Y}}_n\in[0,1]^{T\times L}$ is the prediction tensor, $\mbf{Y}_n\in\{0,1\}^{T\times L}$ is the target tensor, $\epsilon>0$ is a small stabilizing constant, $T$ is the number of label segments, and $L$ is the number of target sound classes. 

In this paper, we experiment with the BCE-Dice loss, which is a combination of the BCE and Dice losses, given by
\begin{equation}
    \mathcal{L}_\text{BCE-Dice}(\hat{\mbf{Y}}, \mbf{Y})
    = \mathcal{L}_\text{BCE}(\hat{\mbf{Y}}, \mbf{Y}) + \mathcal{L}_\text{Dice}(\hat{\mbf{Y}}, \mbf{Y}).
\end{equation}

\subsection{Network architecture}
\label{ssec:archi}

The general architecture of the SED network used in this paper is that of a CRNN, a \textit{de facto} standard architecture for SED~\cite{cakir2017convolutional, Mesaros2021SoundTutorial}. The input is passed through a CNN backbone and pooled in the frequency axis before being passed into an RNN and a fully-connected (FC) output layer.

In this paper, we experiment with a number of different CNN backbones from both audio and vision domains, such as Pretrained Audio Neural Networks (PANNs)~\cite{kong2019panns}, ResNet~\cite{resnet}, EfficientNet~\cite{Tan2019EfficientNet:Networks}, DenseNet~\cite{Huang2016DenselyNetworks}. For ease of comparison, we only use a single-layer 256-unit bidirectional gated recurrent unit (BiGRU) as the RNN network. PyTorch Image Models implementations~\cite{rw2019timm} were adapted for all CNN backbones except PANNs. 

\subsection{Transfer learning}

Early research works into the applications of deep learning in the audio domain suffered severely from the lack of data. 
Many early audio systems relied on the AudioSet-pretrained VGGish model~\cite{Hershey2017CNNClassification, Gemmeke2017AudioEvents}. The use of intra-modal (audio-for-audio) transfer learning remains popular as seen by the frequent use of PANNs~\cite{kong2019panns} in many audio systems today. Recently, more works have successfully attempted inter-modal (vision-for-audio) transfer learning by fine-tuning networks with CNN backbones pretrained on ImageNet~\cite{Kawaguchi2021DescriptionConditions, Giri2020UnsupervisedEstimation, Amiriparian2020TowardsNetworks, Deng2010ImageNet:Database}. Unfortunately, research into the effect of modalities on transfer learning for audio systems remains very limited~\cite{Koike2020AudioClassification}.

In this paper, we investigate the effect of modalities in transfer learning for SED, using CNN backbones pretrained on either single-channel AudioSet~\cite{Gemmeke2017AudioEvents} or three-channel ImageNet~\cite{Deng2010ImageNet:Database}. Since SSE21 provides four-channel inputs, the weight replication technique from~\cite{rw2019timm} is used on the first convolutional layer in order to match the original pretrained weights to the four-channel inputs. 

\subsection{Training procedures}

Recordings from the training set are chunked into four-second segments with a hop size of \SI{0.5}{\second} unless otherwise stated. All models are trained for 30 epochs with a batch size of 32. We use the Adam optimizer~\cite{Kingma2014Adam:Optimization} with a learning rate schedule shown in \Cref{fig:lr}. 

\begin{figure}[t]
    \centering
    \begin{tikzpicture}
    \begin{axis}[
        xmin=-5, xmax=105,
        ymin=-4.5, ymax=-2.5,
        xtick={0, 10, 70, 100},
        ytick={-3, -4},
        yticklabels={\num{e-3}, \num{e-4}},
        axis lines=left,
        height=2.5cm,
        width=0.8\columnwidth,
        grid,
        ylabel=rate,
        ylabel near ticks,
        xlabel near ticks
    ]
    \draw[blue,fill=blue] (axis cs:0,-4) circle (1pt) -- (axis cs:10,-3) circle (1pt) -- (axis cs:70,-3) circle (1pt) -- (axis cs:100,-4) circle (1pt);
    \end{axis}
    \end{tikzpicture}
    \vspace{-1em}
    \caption{Learning rate schedule w.r.t. training epoch (\%)}
    \label{fig:lr}
\end{figure}
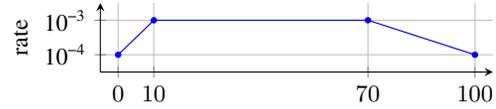

\subsection{Evaluation metrics}

We evaluate the SED performance using the segment-based \Fl score and error rate (ER) with a segment length of \SI{1}{\second}, which have become the standard metrics for SED~\cite{Mesaros2016_MDPI, Mesaros2021SoundTutorial}. 
We combine the ER and \Fl score into a single metric for checkpoint selection as follows: $\text{SEDE} = 0.5 \cdot\text{ER} + 0.5\cdot(1 - \text{\Fl})$. A lower SEDE generally indicates better model performance. All reported results were evaluated on the test set with a binarization threshold of \num{0.3}.

\section{Experiments, Results, and Discussion}
\label{sec:exp}

\subsection{Experiment 1: Data augmentation}

The first experiment investigates the effect of data augmentation techniques on the MSED performance. All models in this experiment used PANN ResNet22 backbone and were trained from scratch with the BCE loss. All 16 combinations of the four augmentation techniques in \Cref{ssec:aug} were tested. All techniques except mixup were each applied with an independent activation probability of \num{0.5}. Mixup used an activation probability of \num{0.8}. 

Mixup (MU) uses soft label mixing with the weight drawn from $\text{Beta}(0.5, 0.5)$ and no mixing is applied if the weight lies in $[0.3, 0.7]$. Frequency shifting (FS) randomly shifts the spectrogram up or down by up to \num{10} mel bins. Channel swapping (CS) randomly selects one of the 24 permutations of the four channels.

The composite cutout-SpecAugment technique (CO) goes as follows. First, one of the these three techniques is chosen at random: single cutout, multiple cutouts, and SpecAugment. Single cutout fills a same-ratio rectangular section randomly located on the spectrogram, sized between \SI{2}{\percent} and \SI{30}{\percent} of the total area, with a constant value drawn randomly from the support of the original input. Multiple cutouts is similar but with \num{8} randomly located sections of size \num{8} bins by \num{8} frames each. SpecAugment uses a time stripe with a width up to \SI{15}{\percent} of the frames and a frequency strip with a width up to \SI{20}{\percent} of the mel bins.

The results are shown in \Cref{tab:daug}. For the FOA format, the ER is the lowest with all four augmentation techniques applied, while the \Fl score is the highest when all except mixup was applied, although the \Fl score for the all-four setting is very close to the no-mixup setting. For the MIC format, the best performance is achieved using the combination of all techniques except mixup. In fact, adding mixup seems to generally worsen the performance compared to the same combination without mixup. FS seems particular useful for both format. CS is useful for FOA format while CO is useful for MIC format. Majority of methods work better in combination than standalone. Interestingly, with or without data augmentation, MIC format provides better performance than FOA format. 

We suspect that the difference in results is due to the difference in how spatial information is stored in the FOA and MIC formats. All channels of MIC formats tend to have similar magnitude information since spatial information is mostly encoded in the phase. The drop in performance when channel swapping was applied on MIC format supports this; we suspect swapping in the MIC format introduces little to no new information in the augmented samples. Moreover, sound event overlaps on spectrograms in the MIC format are often irresolvable without phase information. Hence, mixup likely made the overlapping sound events too difficult to detect by the network. On the other hand, the FOA format encodes spatial information via the relative intensity across the channels with little spatial information in the phase. As such, when mixup is applied, sound events originating from distinct directions are more likely to be resolvable, adding variety to the sound samples and allowing better detection performance. For the subsequent experiments, all four data augmentation techniques were used for FOA format while all except mixup was applied for MIC format.

\begin{table}[t]
    \centering
    \footnotesize
    \begin{tabularx}{\columnwidth}{
        *{4}{C}
        *{4}{S[
            detect-weight, 
            mode=text, 
            table-format=1.3]
        }
    }
    \toprule
        \multicolumn{4}{c}{Data Augmentation} &  
        \multicolumn{2}{c}{FOA} &
        \multicolumn{2}{c}{MIC}\\
        \cmidrule(lr){1-4}\cmidrule(lr){5-6}\cmidrule(lr){7-8}
        MU &  
        CO &  
        FS &
        CS &
        $\text{ER} \downarrow$ &
        $\text{F\textsubscript{1}} \uparrow$ &
        $\text{ER} \downarrow$ &
        $\text{F\textsubscript{1}} \uparrow$ \\
    \midrule
        \V & \V & \V & \V &
            0.459 & 0.670 & 
            0.420 & 0.714 \\\midrule
        \X & \V & \V & \V &
            0.452 & 0.681 &
            0.438 & 0.697 \\
        \V & \X & \V & \V & 
            0.464 & 0.665 & 
            0.398 & \emph{0.725} \\
        \V & \V & \X & \V & 
            \emph{0.427} & \emph{0.694} & 
            \emph{0.395} & 0.721 \\
        \V & \V & \V & \X &
            0.446 & 0.679 &
            0.458 & 0.675 \\ \midrule
        \X & \X & \V & \V &
            0.449 & 0.689 &
            0.445 & 0.683 \\
        \X & \V & \X & \V & 
            0.424 & 0.695 & 
            0.404 & 0.723 \\
        \X & \V & \V & \X & 
            0.471 & 0.658 &
            0.479 & 0.649 \\
        \V & \X & \X & \V & 
            0.404 & 0.720 & 
            \emph{0.344} & \emph{0.775} \\
        \V & \X & \V & \X & 
            0.470 & 0.658 & 
            0.458 & 0.666 \\
        \V & \V & \X & \X & 
            \emph{0.389} & \emph{0.737} & 
            0.413 & 0.702\\\midrule
        \X & \X & \X & \V &
            0.410 & 0.700 &
            0.432 & 0.687 \\
        \X & \X & \V & \X & 
            0.434 & 0.684 & 
            0.432 & 0.687\\ 
        \X & \V & \X & \X &
            0.434 & 0.684 &
            0.405 & 0.707\\ 
        \V & \X & \X & \X & 
            \emph{0.390} & \B 0.736 & 
            \B 0.339 & \B 0.775\\\midrule
        \X & \X & \X & \X &
            \B 0.384 & 0.734 &
            0.424 & 0.698 \\
    \bottomrule`
    \end{tabularx}
    \caption{SED performance w.r.t. data augmentation.}
    \label{tab:daug}
\end{table}

\subsection{Experiment 2: Loss function}

\begin{table}[t]
    \footnotesize
    \centering
    \begin{tabularx}{\columnwidth}{
        Xc
        *{4}{S[
            detect-weight, 
            mode=text, 
            table-format=1.3]
        }
    }
    \toprule
        &&
        \multicolumn{2}{c}{FOA} &
        \multicolumn{2}{c}{MIC} \\
        \cmidrule(lr){3-4}\cmidrule(lr){5-6}
        Loss function&  
        Transfer&
        $\text{ER} \ \downarrow$ &
        $\text{F\textsubscript{1}} \uparrow$ &
        $\text{ER} \downarrow$ &
        $\text{F\textsubscript{1}} \uparrow$ \\
    \midrule
        BCE & \V & 
            0.384 & 0.734 &
            0.339 & 0.775  \\ 
        BCE & \X & 
            0.371 & 0.739 &
            0.353 & 0.760 \\
        BCE + Dice & \V &
            0.372 & 0.744 &
            0.377 & 0.728 \\
        BCE + Dice & \X &
            \B 0.337 & \B 0.762 &
            \B 0.332 & \B 0.779 \\
    \bottomrule`
    \end{tabularx}
    \caption{SED performance w.r.t. loss function and transfer learning.}
    \label{tab:dice}
\end{table}

The second experiment investigates the effect of loss functions on the performance of SED models with and without transfer learning. 
The experiment settings were identical to that of Experiment~1. 

The results for this experiment are shown in \Cref{tab:dice}. For both formats, the best performance was achieved with both transfer learning and the BCE-Dice loss function. 
This result is somewhat expected as the Dice loss is generally more robust to positive-negative imbalance in the dataset than the BCE loss, and can be considered a soft approximation of the binary \Fl score~\cite{Li2020DiceTasks}. Interestingly, using only either the BCE-Dice loss or pretraining worsened the performance in the MIC format, but using both improved the performance.
This is not observed in the FOA format, where using either technique improved the performance, and using both improved the performance further.

\subsection{Experiment 3: Training chunk size}

\begin{table}[t]
    \footnotesize
    \centering
    \begin{tabular}{
        S[
            detect-weight, 
            mode=text, 
            table-format=1.1]
        *{4}{S[
            detect-weight, 
            mode=text, 
            table-format=1.3]
        }
    }
    \toprule
        &
        \multicolumn{2}{c}{FOA} &
        \multicolumn{2}{c}{MIC} \\
        \cmidrule(lr){2-3}\cmidrule(lr){4-5}
        \multicolumn{1}{l}{Chunk size (\si{\second})}&
        $\text{ER} \ \downarrow$ &
        $\text{F\textsubscript{1}} \uparrow$ &
        $\text{ER} \downarrow$ &
        $\text{F\textsubscript{1}} \uparrow$ \\
    \midrule
        4.0 & 0.337 & 0.762 & 0.332 & \B 0.779\\
        8.0 & 0.339 & 0.764 & 0.349 & 0.765\\
        12.0 & \B 0.318 & \B 0.792 & \B 0.326 & 0.770\\
    \bottomrule`
    \end{tabular}
    \caption{SED performance w.r.t. training chunk size.}
    \label{tab:chunk}
\end{table}

The third experiment investigates the impact of the training chunk size on the SED performance. For reference, event lengths for SSE21 have a median of \SI{3.2}{\second} and a mean of \SI{8.3}{\second}. We used both transfer learning and the BCE-Dice loss for training. The rest of the settings followed Experiment~2.

The results for this experiment are shown in \Cref{tab:chunk}. For the FOA format, increasing the chunk size from \SI{4}{\second} to \SI{8}{\second} barely improved the performance, but increasing to \SI{12}{\second} resulted in a some appreciable gain. The results for the MIC format are less consistent, with the \SI{8}{\second} chunk performing the worst. Further analysis of SSE21 revealed some sound classes with a significant number of events whose durations are beyond \SI{8}{\second}. Insufficient coverage by the \SI{8}{\second} chunk may explain the lack in performance gain, as training difficulty increased beyond the additional information gained.

\subsection{Experiment 4: Input channels}

The fourth experiment compares the performance of SED networks using single-channel inputs against those using the full four-channel array inputs. For the single-channel setting, the omnidirectional (W) channel was used for the FOA format while the first channel was used for the MIC format. A chunk size of \SI{4}{s} was used and the rest of the settings followed Experiment~3.

The results are shown in \Cref{tab:chan}. With the addition of more channels, the FOA format achieved very little performance gain, if any, whereas the MIC format gained significant improvement. The similar performances for the FOA format are likely because the information in the W channel was already acquired indirectly via multiple microphones in an ambisonic array (32 sensors for SSE21). On the other hands, each channel in the MIC format only acquired information from a single sensor (which is also more representative of a single-sensor signal acquisition in practice). As such, the information gained from adding more channels in the MIC format is more significant, compared to the mostly spatial information gained from more channels in the FOA format. 

\begin{table}[t]
    \footnotesize
    \centering
    \begin{tabular}{
        l
        *{4}{S[
            detect-weight, 
            mode=text, 
            table-format=1.3]
        }
    }
    \toprule
        &
        \multicolumn{2}{c}{FOA} &
        \multicolumn{2}{c}{MIC} \\
        \cmidrule(lr){2-3}\cmidrule(lr){4-5}
        Channels&
        $\text{ER} \ \downarrow$ &
        $\text{F\textsubscript{1}} \uparrow$ &
        $\text{ER} \downarrow$ &
        $\text{F\textsubscript{1}} \uparrow$ \\
    \midrule
        Mono    & 0.340 & \B 0.763 & 0.375 & 0.740\\
        All     & \B 0.337 & 0.762 & \B 0.332 & \B 0.779\\
    \bottomrule`
    \end{tabular}
    \caption{SED performance w.r.t. number of channels.}
    \label{tab:chan}
\end{table}

\subsection{Experiment 5: Modalities in transfer learning}

\begin{table}[t]
    \footnotesize
    \centering
    \begin{tabularx}{\columnwidth}{
        Xcl
        *{4}{S[
            detect-weight, 
            mode=text, 
            table-format=1.3]
        }
    }
    \toprule
        &&&
        \multicolumn{2}{c}{FOA} &
        \multicolumn{2}{c}{MIC} \\
        \cmidrule(lr){4-5}\cmidrule(lr){6-7}
        CNN&
        DF &
        TL&
        $\text{ER} \ \downarrow$ &
        $\text{F\textsubscript{1}} \uparrow$ &
        $\text{ER} \downarrow$ &
        $\text{F\textsubscript{1}} \uparrow$ \\
    \midrule
        PANN ResNet22        & 16    & Audio 
            & \B 0.337  & \B 0.762 
            & \B 0.332  & \B 0.779\\
        PANN CNN14      & 16    & Audio 
            & 0.394     & 0.739
            & 0.357	    & 0.767 \\
        ResNet18*       & 16    & None  
            & 0.371     & 0.744 
            & 0.378     & 0.749 \\
        PANN ResNet22        & 16    & None  
            & 0.372     & 0.744 
            & 0.377     & 0.728 \\
        EfficientNetB0  & 32    & Image 
            & 0.398     & 0.722 
            & 0.376     & 0.743 \\
        PANN CNN14      & 16    & None  
            & 0.446     & 0.685 
            & 0.365	    & 0.760 \\
        ResNet18*       & 16    & Image 
            & 0.437     & 0.680 
            & 0.365     & 0.743 \\
        ResNet18        & 32    & None  
            & 0.430     & 0.701
            & 0.415     & 0.720 \\
        DenseNet121     & 32    & Image 
            & 0.434     & 0.688
            & 0.421     & 0.731 \\
        ResNet18        & 32    & Image 
            & 0.423     & 0.706
            & 0.423     & 0.698 \\
        EfficientNetB0  & 32    & None  
            & 0.463     & 0.673 
            & 0.454     & 0.692 \\
        DenseNet121     & 32    & None  
            & 0.474     & 0.650 
            & 0.433     & 0.689 \\
    \bottomrule
    \end{tabularx}
    \caption{SED performance w.r.t. the CNN backbone and transfer learning, listed in decreasing average SEDE. The ResNet18* model was modified from ResNet18 by removing the first pooling layer.}
    \label{tab:cnn}
\end{table}

In the last experiment, we investigate the impact of pretraining modalities in transfer learning (TL) on the MSED performance with various CNN backbones. The settings followed Experiment~4 with all channels as inputs.

The results for this experiment are shown in \Cref{tab:cnn}. For both formats, the pretrained networks generally perform better than their counterparts trained from scratch. This confirms the conventional wisdom that transfer learning, from the same or a related domain, allows deep networks to utilize some previously learnt patterns which are useful to SED. The ResNet18* backbone had some interesting results: With ResNet18*, where the original weights meant for ResNet18 were used, pretraining worsened performance in the FOA format but improved performance in the MIC format.

In terms of architecture, it is clear that models with ResNet-based or CNN14 backbones outperformed those with EfficientNet or DenseNet backbone when trained from scratch. DenseNet architecture is a known to cause aliasing problems \cite{Takahashi2021} which may explain the poor performance. EfficientNet has building blocks which consist of information compressing layers, potentially causing in compounding information loss when trained on small datasets. 

In terms of modalities, comparing the PANNs pretrained on AudioSet to other CNNs pretrained on ImageNet, it is clear networks with AudioSet-pretrained backbones outperformed those with ImageNet-pretrained backbones. This is consistent with the findings from~\cite{Koike2020AudioClassification} where intra-modal transfer learning outperformed inter-modal ones. Since higher-level features in audio spectrograms are very different from higher-level features in images, we suspect that higher-level representation prelearning of audio-related patterns may only be achieved via intra-modal pretraining, even if pretraining on vision datasets gives the network a headstart in learning lower-level visual representation.

\section{CONCLUSION}
\label{sec:concl}

In this paper, we performed a thorough investigation on the factors affecting the performance of deep multichannel polyphonic SED systems: data augmentation techniques, loss functions, training chunk durations, pretraining modalities, and multichannel audio formats. A proper combination of data augmentation techniques can mitigate the problem of small datasets, and significantly improve the model performance. Further improvement can be achieved by using the BCE-Dice loss and transfer learning. We showed that both inter- and intra-model transfer learning from other datasets increases the SED performance on multi-channel datasets, even with channel mismatch, with intra-model transfer learning providing higher performance gains. Interestingly, across many settings, the best performances of MIC format are slightly higher than those of FOA format.

\newcommand{\fts}{9pt}
\newcommand{\sks}{9.2pt}
\renewcommand{\bibliosize}{\fontsize{\fts}{\sks}\selectfont}
{
    \bibliographystyle{IEEEtran}
    \bibliography{refs}

\begin{thebibliography}{10}
\providecommand{\url}[1]{#1}
\def\UrlFont{\rmfamily}
\providecommand{\newblock}{\relax}
\providecommand{\bibinfo}[2]{#2}
\providecommand\BIBentrySTDinterwordspacing{\spaceskip=0pt\relax}
\providecommand\BIBentryALTinterwordstretchfactor{4}
\providecommand\BIBentryALTinterwordspacing{\spaceskip=\fontdimen2\font plus
\BIBentryALTinterwordstretchfactor\fontdimen3\font minus
  \fontdimen4\font\relax}
\providecommand\BIBforeignlanguage[2]{{%
\expandafter\ifx\csname l@#1\endcsname\relax
\typeout{** WARNING: IEEEtran.bst: No hyphenation pattern has been}%
\typeout{** loaded for the language `#1'. Using the pattern for}%
\typeout{** the default language instead.}%
\else
\language=\csname l@#1\endcsname
\fi
#2}}

\bibitem{Mesaros2021SoundTutorial}
\BIBentryALTinterwordspacing
A.~Mesaros, T.~Heittola, T.~Virtanen, and M.~D. Plumbley, ``{Sound Event
  Detection: A Tutorial},'' \emph{\protect\JournalTitle{IEEE Signal Processing
  Magazine}}, vol.~38, no.~5, 2021.
\BIBentrySTDinterwordspacing

\bibitem{kong2019panns}
\BIBentryALTinterwordspacing
Q.~Kong, Y.~Cao, T.~Iqbal, Y.~Wang, W.~Wang, and M.~D. Plumbley, ``{PANNs:
  Large-Scale Pretrained Audio Neural Networks for Audio Pattern
  Recognition},'' \emph{\protect\JournalTitle{IEEE/ACM Transactions on Audio
  Speech and Language Processing}}, vol.~28, pp. 2880--2894, 2020.
\BIBentrySTDinterwordspacing

\bibitem{cakir2017convolutional}
E.~Cakir, G.~Parascandolo, T.~Heittola, H.~Huttunen, and T.~Virtanen,
  ``{Convolutional Recurrent Neural Networks for Polyphonic Sound Event
  Detection},'' \emph{\protect\JournalTitle{IEEE/ACM Transactions on Audio
  Speech and Language Processing}}, vol.~25, no.~6, pp. 1291--1303, 2017.

\bibitem{Miyazaki2020Conformer-BasedAugmentation}
K.~Miyazaki, T.~Komatsu, T.~Hayashi, S.~Watanabe, T.~Toda, K.~Takeda, and
  L.~Corporation, ``{Conformer-Based Sound Event Detection With Semi-Supervised
  Learning and Data Augmentation},'' in \emph{\protect\JournalTitle{Proceedings
  of the 5th Workshop on Detection and Classification of Acoustic Scenes and
  Events}}, 2020, pp. 100--104.

\bibitem{Gemmeke2017AudioEvents}
J.~F. Gemmeke, D.~P. Ellis, D.~Freedman, A.~Jansen, W.~Lawrence, R.~C. Moore,
  M.~Plakal, and M.~Ritter, ``{Audio Set: An ontology and human-labeled dataset
  for audio events},'' in \emph{\protect\JournalTitle{Proceedings of the IEEE
  International Conference on Acoustics, Speech and Signal Processing}}, 2017,
  pp. 776--780.

\bibitem{Fonseca2017FreesoundDatasets}
E.~Fonseca, J.~Pons, X.~Favory, F.~Font, D.~Bogdanov, A.~Ferraro, S.~Oramas,
  A.~Porter, and X.~Serra, ``{Freesound datasets: A platform for the creation
  of open audio datasets},'' in \emph{\protect\JournalTitle{Proceedings of the
  18th International Society for Music Information Retrieval Conference}},
  2017, pp. 486--493.

\bibitem{Mesaros2016TUTDetection}
A.~Mesaros, T.~Heittola, and T.~Virtanen, ``{TUT database for acoustic scene
  classification and sound event detection},'' in
  \emph{\protect\JournalTitle{Proceedings of the European Signal Processing
  Conference}}, 2016, pp. 1128--1132.

\bibitem{politis2021dcasedataset}
A.~Politis, S.~Adavanne, D.~Krause, A.~Deleforge, P.~Srivastava, and
  T.~Virtanen, ``{A Dataset of Dynamic Reverberant Sound Scenes with
  Directional Interferers for Sound Event Localization and Detection},''
  \emph{\protect\JournalTitle{arXiv}}, 2021.

\bibitem{adavanne2017sedspatial}
S.~Adavanne, P.~Pertila, and T.~Virtanen, ``{Sound event detection using
  spatial features and convolutional recurrent neural network},'' in
  \emph{\protect\JournalTitle{Proceedings of the IEEE International Conference
  on Acoustics, Speech and Signal Processing}}, 2017, pp. 771--775.

\bibitem{Nguyen2020OnDetection}
T.~N.~T. Nguyen, D.~L. Jones, and W.~S. Gan, ``{On the Effectiveness of Spatial
  and Multi-Channel Features for Multi-Channel Polyphonic Sound Event
  Detection},'' in \emph{\protect\JournalTitle{Proceedings of the 5th Workshop
  on Detection and Classification of Acoustic Scenes and Events}}, 2020, pp.
  115--119.

\bibitem{Imoto2021ImpactPerformance}
K.~Imoto, S.~Mishima, Y.~Arai, and R.~Kondo, ``{Impact of Sound Duration and
  Inactive Frames on Sound Event Detection Performance},'' in
  \emph{\protect\JournalTitle{Proceedings of the IEEE International Conference
  on Acoustics, Speech and Signal Processing}}, 2021, pp. 860--864.

\bibitem{politis2020dcasedataset}
A.~Politis, S.~Adavanne, and T.~Virtanen, ``{A Dataset of Reverberant Spatial
  Sound Scenes with Moving Sources for Sound Event Localization and
  Detection},'' \emph{\protect\JournalTitle{arXiv}}, 2020.

\bibitem{zhang2017mixup}
\BIBentryALTinterwordspacing
H.~Zhang, M.~Cisse, Y.~N. Dauphin, and D.~Lopez-Paz, ``{MixUp: Beyond empirical
  risk minimization},'' in \emph{\protect\JournalTitle{Conference Track
  Proceedings of the 6th International Conference on Learning
  Representations}}, 2018.

\bibitem{Mazzon2019}
L.~Mazzon, Y.~Koizumi, M.~Yasuda, and N.~Harada, ``{First Order Ambisonics
  Domain Spatial Augmentation for DNN-based Direction of Arrival Estimation},''
  in \emph{\protect\JournalTitle{Proceedings of the 4th Workshop on Detection
  and Classification of Acoustic Scenes and Events}}, 2019, pp. 154--158.

\bibitem{DeVries2017ImprovedCutout}
\BIBentryALTinterwordspacing
T.~DeVries and G.~W. Taylor, ``{Improved Regularization of Convolutional Neural
  Networks with Cutout},'' \emph{\protect\JournalTitle{arXiv}}, 2017.
\BIBentrySTDinterwordspacing

\bibitem{park2019specaugment}
D.~S. Park, W.~Chan, Y.~Zhang, C.~C. Chiu, B.~Zoph, E.~D. Cubuk, and Q.~V. Le,
  ``{SpecAugment: A simple data augmentation method for automatic speech
  recognition},'' in \emph{\protect\JournalTitle{Proceedings of the Annual
  Conference of the International Speech Communication Association}}, 2019, pp.
  2613--2617.

\bibitem{Dice1945MeasuresSpecies}
L.~R. Dice, ``{Measures of the Amount of Ecologic Association Between
  Species},'' \emph{\protect\JournalTitle{Ecology}}, vol.~26, no.~3, pp.
  297--302, 1945.

\bibitem{Srensen1948ACommons}
T.~S{\o}rensen, ``{A method of establishing roups of equal amplitude in plant
  sociology based on similarity of species and its application to analyses of
  the vegetation on Danish commons},'' \emph{\protect\JournalTitle{Kongelige
  Danske Videnskabernes Selskab}}, vol.~5, no.~4, pp. 1--34, 1948.

\bibitem{Milletari2016V-Net:Segmentation}
F.~Milletari, N.~Navab, and S.~A. Ahmadi, ``{V-Net: Fully convolutional neural
  networks for volumetric medical image segmentation},'' in
  \emph{\protect\JournalTitle{Proceedings of the 4th International Conference
  on 3D Vision}}, 2016, pp. 565--571.

\bibitem{Sudre2017GeneralisedSegmentations}
C.~H. Sudre, W.~Li, T.~Vercauteren, S.~Ourselin, and M.~J. Cardoso,
  ``{Generalised Dice overlap as a deep learning loss function for highly
  unbalanced segmentations},'' in \emph{\protect\JournalTitle{Deep Learning in
  Medical Image Analysis and Multimodal Learning for Clinical Decision
  Support}}.\hskip 1em plus 0.5em minus 0.4em\relax Springer, 2017, pp.
  240--248.

\bibitem{resnet}
\BIBentryALTinterwordspacing
K.~He, X.~Zhang, S.~Ren, and J.~Sun, ``{Deep residual learning for image
  recognition},'' in \emph{\protect\JournalTitle{Proceedings of the IEEE
  Computer Society Conference on Computer Vision and Pattern Recognition}},
  2016, pp. 770--778.

\bibitem{Tan2019EfficientNet:Networks}
M.~Tan and Q.~V. Le, ``{EfficientNet: Rethinking model scaling for
  convolutional neural networks},'' in \emph{\protect\JournalTitle{Proceedings
  of the 36th International Conference on Machine Learning}}, 2019, pp.
  10\,691--10\,700.

\bibitem{Huang2016DenselyNetworks}
\BIBentryALTinterwordspacing
G.~Huang, Z.~Liu, L.~van~der Maaten, and K.~Q. Weinberger, ``{Densely Connected
  Convolutional Networks},'' in \emph{\protect\JournalTitle{Proceedings of the
  30th IEEE Conference on Computer Vision and Pattern Recognition}}, 2016, pp.
  2261--2269.

\bibitem{rw2019timm}
R.~Wightman, ``{PyTorch Image Models},'' 2019.

\bibitem{Hershey2017CNNClassification}
\BIBentryALTinterwordspacing
S.~Hershey, S.~Chaudhuri, D.~P. Ellis, J.~F. Gemmeke, A.~Jansen, R.~C. Moore,
  M.~Plakal, D.~Platt, R.~A. Saurous, B.~Seybold, M.~Slaney, R.~J. Weiss, and
  K.~Wilson, ``{CNN architectures for large-scale audio classification},'' in
  \emph{\protect\JournalTitle{Proceedings of the IEEE International Conference
  on Acoustics, Speech and Signal Processing}}, 2017, pp. 131--135.

\bibitem{Kawaguchi2021DescriptionConditions}
\BIBentryALTinterwordspacing
Y.~Kawaguchi, K.~Imoto, Y.~Koizumi, N.~Harada, D.~Niizumi, K.~Dohi, R.~Tanabe,
  H.~Purohit, and T.~Endo, ``{Description and Discussion on DCASE 2021
  Challenge Task 2: Unsupervised Anomalous Sound Detection for Machine
  Condition Monitoring under Domain Shifted Conditions},''
  \emph{\protect\JournalTitle{arXiv}}, 2021.
\BIBentrySTDinterwordspacing

\bibitem{Giri2020UnsupervisedEstimation}
R.~Giri, S.~V. Tenneti, F.~Cheng, and K.~Helwani, ``{Unsupervised Anomalous
  Sound Detection Using Self-Supervised Classification and Group Masked
  Autoencoder for Density Estimation},'' \emph{\protect\JournalTitle{IEEE AASP
  Challenge on Detection and Classification of Acoustic Scenes and Events
  2020}}, no.~2, 2020.

\bibitem{Amiriparian2020TowardsNetworks}
\BIBentryALTinterwordspacing
S.~Amiriparian, M.~Gerczuk, S.~Ottl, L.~Stappen, A.~Baird, L.~Koebe, and
  B.~Schuller, ``{Towards cross-modal pre-training and learning tempo-spatial
  characteristics for audio recognition with convolutional and recurrent neural
  networks},'' \emph{\protect\JournalTitle{Eurasip Journal on Audio, Speech,
  and Music Processing}}, vol. 2020, no.~1, pp. 1--11, 2020.
\BIBentrySTDinterwordspacing

\bibitem{Deng2010ImageNet:Database}
J.~Deng, W.~Dong, R.~Socher, L.-J. Li, {Kai Li}, and {Li Fei-Fei}, ``{ImageNet:
  A large-scale hierarchical image database},'' in
  \emph{\protect\JournalTitle{Proceedings of the IEEE Conference on Computer
  Vision and Pattern Recognition}}, 2010, pp. 248--255.

\bibitem{Koike2020AudioClassification}
T.~Koike, K.~Qian, Q.~Kong, M.~D. Plumbley, B.~W. Schuller, and Y.~Yamamoto,
  ``{Audio for Audio is Better? An Investigation on Transfer Learning Models
  for Heart Sound Classification},'' in \emph{\protect\JournalTitle{Proceedings
  of the Annual International Conference of the IEEE Engineering in Medicine
  and Biology Society}}, 2020, pp. 74--77.

\bibitem{Kingma2014Adam:Optimization}
\BIBentryALTinterwordspacing
D.~P. Kingma and J.~Ba, ``{Adam: A Method for Stochastic Optimization},''
  \emph{\protect\JournalTitle{Conference Track Proceedings of the 3rd
  International Conference on Learning Representations}}, 2014.
\BIBentrySTDinterwordspacing

\bibitem{Mesaros2016_MDPI}
A.~Mesaros, T.~Heittola, and T.~Virtanen, ``{Metrics for polyphonic sound event
  detection},'' \emph{\protect\JournalTitle{Applied Sciences (Switzerland)}},
  vol.~6, no.~6, p. 162, 2016.

\bibitem{Li2020DiceTasks}
\BIBentryALTinterwordspacing
X.~Li, X.~Sun, Y.~Meng, J.~Liang, F.~Wu, and J.~Li, ``{Dice Loss for
  Data-imbalanced NLP Tasks},'' in \emph{\protect\JournalTitle{Proceedings of
  the 58th Annual Meeting of the Association for Computational Linguistics}},
  2020, pp. 465--476.

\bibitem{Takahashi2021}
\BIBentryALTinterwordspacing
N.~Takahashi and Y.~Mitsufuji, ``{Densely connected multidilated convolutional
  networks for dense prediction tasks},'' in
  \emph{\protect\JournalTitle{Proceedings of the IEEE/CVF Conference on
  Computer Vision and Pattern Recognition}}, 2021.

\end{thebibliography}
}

\end{sloppy}
\end{document}